# Predicting the Thermal Conductivity Collapse in SWCNT Bundles: The Interplay of Symmetry Breaking and Scattering Revealed by Machine-Learning-Driven Quantum Transport

*Feng Tao[1], Xiaoliang Zhang[1], Dawei Tang[1], Shigeo Maruyama[2,3,4\*], Ya Feng[1,3\*]*

[1]Key Laboratory of Ocean Energy Utilization and Energy Conservation of Ministry of Education, School of Energy and Power Engineering, Dalian University of Technology, Dalian 116024, China

[2]State Key Laboratory of Fluid Power and Mechatronic Systems, School of Mechanical Engineering, Zhejiang University, Hangzhou 310027, China

[3]Department of Mechanical Engineering, The University of Tokyo, Tokyo 113-8656, Japan

[4]Institute of Materials Innovation, Institutes for Future Society, Nagoya University, Nagoya 464-8603, Japan

**ABSTRACT:** We combine machine learning (ML)-based neuroevolution potentials (NEP) with anharmonic lattice dynamics and the Boltzmann transport equation (ALD-BTE) to achieve a quantitative and mode-resolved description of thermal transport in individual (10, 0) zigzag single-walled carbon nanotubes (SWCNTs) and their bundles. Our analysis reveals a dual mechanism behind the drastic suppression of thermal conductivity in bundles: first, the breaking of rotational symmetry in isolated SWCNTs dramatically enhances the scattering rates of symmetry-sensitive phonon modes, such as the twist (TW) mode. Second, the emergence of new inter-tube phonon modes introduces abundant additional scattering channels across the entire frequency spectrum. Crucially, the incorporation of quantum Bose-Einstein (BE) statistics is essential to accurately capture these phenomena, enabling our approach to quantitatively reproduce experimental observations. This work establishes the combination of ML-driven interatomic potentials and ALD-BTE as a predictive framework for nanoscale thermal transport, effectively bridging the gap between theoretical models and experimental measurements.

**KEYWORDS:** SWCNTs; bundles; machine learning potential; Bose-Einstein statistics; phonon transport

# INTRODUCTION

Single-walled carbon nanotubes (SWCNTs) have long been considered ideal candidates for thermal management applications, such as phase-change materials, thermal interfaces, and electronic cooling, due to their exceptionally high thermal conductivity.[1, 2] However, in practical macroscopic assemblies, SWCNTs inevitably form bundles through van der Waals (vdW) interactions, which drastically suppress thermal transport. Experiments consistently report severe reductions in thermal conductivity upon bundling, independent of chirality:[2-7] for instance, a 5-μm-long bundle of three identical SWCNTs shows a 75 % decrease compared to its isolated counterpart,[2] while bundles with mixed chiralities exhibit reductions of 70 % (four-SWCNT) and 97 % (thirteen-SWCNT) in contrast to the individual one.[5] This effect arises primarily from inter-tube vdW interactions, which intensify phonon scattering between adjacent SWCNTs.[4, 8, 9] Despite these experimental observations, theoretical studies struggle to replicate such dramatic reductions. Computational limitations, including imperfect interatomic potentials and resource constraints, have restricted simulations to oversimplified models, such as the parallel circuit approximation, which fails to capture the experimental trends.[10, 11] Recent molecular dynamics (MD) simulations have made progress, revealing a 10 % thermal conductivity reduction in 10-μm-long bundles due to vdW-induced low-frequency phonon hybridization, and underscoring the inadequacy of the parallel circuit model for extended systems.[12] While this represents a significant advance, a comprehensive theoretical framework bridging simulations and experiments remains elusive.

The curvature-induced structural complexity of SWCNTs poses significant challenges for empirical potentials, often leading to discrepancies between simulations and experiments as mentioned above. While traditional potentials like Tersoff and AIREBO, accurately describe covalent C–C bonds in planar graphene, they fail to capture interatomic interactions in curved

SWCNT geometries, leading to unreliable predictions of physical properties.[13] Similarly, the widely used Lennard-Jones (L-J) potential substantially underestimates the corrugation of the potential energy surface governing vdW interactions in $sp^2$-bonded carbon systems, including graphite, few-layer graphene, and $C_{60}$–SWCNT hybrids.[14-16] Recent advances in machine learning (ML) potentials offer a promising solution by combining density functional theory (DFT)-level precision with the computational efficiency of MD.[17, 18] For instance, Li *et al*. demonstrated that ML potentials can successfully reproduce the pressure-induced transformation of carbon peapods into novel carbon structures, in agreement with experiments.[19] Beyond potential inaccuracies, the fundamental limitations of classical MD for SWCNTs must be considered. The high Debye temperature (~2500 K) and low-dimensionality of SWCNTs require quantum, namely Bose-Einstein (BE), statistics to describe phonon populations accurately, even at room temperature, a regime where classical equipartition (EQ) fail.[20-22] Furthermore, while MD in the microcanonical (NVE) ensemble conserves energy, it implicitly samples phonon scattering within a classical framework. This approach obscures the distinction between well-defined quantum processes such as phonon absorption/emission and Normal versus Umklapp processes.[23, 24] In contrast, the anharmonic lattice dynamics and the Boltzmann transport equation (ALD-BTE) framework naturally incorporates BE statistics and explicit energy- and momentum-conserving selection rules for three-phonon processes, thereby overcoming the limitations of classical MD.[25] Although ALD-BTE, being a perturbative approach, may fail in strongly anharmonic systems where higher-order scattering dominates,[26] its application to SWCNTs is well-justified. The observed long phonon mean free path in SWCNTs reflect relatively weak anharmonicity, confirming the reliability of the ALD-BTE framework for these systems.[27, 28]

In this work, we develop a high-accuracy neuroevolution potential (NEP) for SWCNTs and integrate it with ALD-BTE framework to achieve a quantitative, mode-resolved understanding of thermal transport in individual (10, 0) zigzag SWCNTs and their bundled assemblies. Our approach accurately reproduces experimental observations, predicting an 81 % reduction in thermal conductivity for a 5-μm-long bundle of seven SWCNTs, a striking effect arising from inter-tube interactions. Through detailed phonon-mode analysis, we demonstrate that bundling breaks the rotational symmetry of SWCNTs, enhances the three-phonon scattering phase space, and elevates phonon scattering rates across the entire phonon spectrum, collectively leading to the dramatic suppression of thermal conductivity. Crucially, replacing BE statistics with EQ statistics eliminates the agreement with experiment, underscoring the essential role of quantum statistics in accurately simulating these systems. Our work highlights the power of ML-driven interatomic potentials combined with ALD-BTE for precise predictions of nanoscale thermal transport, offering a unified framework that bridges theoretical models and experimental measurements. The methodology developed here is generalizable to other low-dimensional materials, while our findings provide key design principles for SWCNT-based thermal management systems, from nanoscale devices to macroscopic applications.

## RESULTS AND DISCUSSION

We calculated the length-dependent thermal conductivity of the (10, 0) SWCNT using both spectral heat current (SHC) and nonequilibrium molecular dynamics (NEMD) with two interatomic potentials: our newly trained NEP and the established Tersoff potential. As shown in Figure 1a, the NEP-based SHC results ($\kappa(L)_{\text{SHC-NEP}}$) are in excellent agreement with our NEMD-NEP calculations, validating the accuracy of the SHC methodology in SWCNT. This agreement is further corroborated by comparing SHC and NEMD results using the Tersoff potential; our $\kappa(L)_{\text{SHC-Tersoff}}$ values show remarkable consistency with prior NEMD-Tersoff data from Donadio et al.[28] across lengths spanning 10 nm to 10 μm (see Section S2.2, Supporting Information (SI)). Notably, $\kappa(L)_{\text{SHC-NEP}}$ is systematically lower than $\kappa(L)_{\text{SHC-Tersoff}}$. We attribute this discrepancy to the Tersoff potential's underestimation of the SWCNT potential energy (~1350 meV/atom), as detailed in Section S3 of SI. This inaccuracy artificially inflates phonon group velocities and suppresses anharmonic scattering, leading to an overestimation of thermal conductivity.

We refer to a bundle consisting of *n* identical SWCNTs as an N-SWCNT bundle and examined cases where, *n* = 1, 2, 3, 5, and 7; all simulations are performed at 300K. For clarity, Figure 1a only presents the $\kappa(L)$ for an individual (1-)SWCNT (red) and 7-SWCNT bundle (blue), calculated using the SHC method and NEMD simulations with the NEP. Corresponding data for all five bundle sizes are provided in Figure S5. Two methods exhibit excellent agreement for effective thermal conductivity at lengths of 20 and 50 nm. The thermal conductivity in the ballistic limit (dashed lines in Figure 1a) was extracted by fitting the SHC results at very short lengths ($L < 2$ nm). In this regime, $\kappa_{\text{SHC}}$ increases linearly with $L$, and the value for the bundles is nearly identical to that of a single SWCNT (1-SWCNT), indicating that intertube coupling has a negligible effect on ballistic transport. Below 1 μm, the $\kappa_{\text{SHC}}$ of the bundles remains comparable to that of the 1-SWCNT and increases rapidly with length. However, for lengths exceeding 1 μm, the $\kappa_{\text{SHC}}$ of the bundles begins to deviate from that of 1-SWCNT, with larger bundles exhibiting progressively lower $\kappa_{\text{SHC}}$. This trend is consistent with prior SHC-based studies by Shiga et al. using the Tersoff potential,[12] suggesting that the variation of thermal conductivity in MD simulations for bundles is largely independent of the interatomic potential employed. The observed reduction in $\kappa$ for longer bundles (> 1 μm) can be attributed to phonon mode hybridization and coupling, driven primarily by vdW interactions that predominantly affect low-frequency phonon modes.[12] Since these modes contribute significantly to thermal transport at larger lengths, their suppression leads to a notable decline in thermal conductivity. This effect is more explicitly illustrated in Figure 1c,d. At a length of $L = 20$ nm, both $\kappa_{NEMD}$ and $\kappa_{SHC}$ show good agreement and exhibit minimal dependent on bundle size (Figure 1c), and the large uncertainty in the $\kappa_{SHC}$ values arises from the ensemble averaging over independent simulation runs.[29] In contrast, at $L = 10$ μm, $\kappa_{SHC}$ continuously decreases with the increasing number of SWCNT in the bundles, with the 7-SWCNT bundle showing 35.4 % decrease compared to 1-SWCNT. Despite the improved accuracy of the NEP, the SHC results from our simulations remains inconsistent with the experimental observations, which report thermal conductivity reductions of 75 % for a 3-SWCNT bundle and 86 % for an 8-SWCNT bundle at a length of 5 μm.

Alternatively, the ALD-BTE framework, incorporates BE statistics and explicitly enforces the energy-conservation selection rule for three-phonon scattering via Fermi's golden rule, offers a more accurate description of thermal transport in SWCNTs.[28, 30] For length-dependent ALD-BTE simulations, we adopt the consistent boundary conditions proposed by Maassen and Lundstrom,[31] which discard the local thermal equilibrium assumption and treat forward and backward fluxes separately along the transport direction. In one-dimensional systems, this approach yields an explicit expression for finite-length thermal conductivity,

$$\kappa(L)_{\text{ALD-BTE}} = \frac{1}{N_q V} \sum_{\mu\mu'} C_\mu v_\mu (\Gamma_{\mu\mu'} + \frac{2|v_{\mu'}|}{L} \delta_{\mu\mu'})^{-1} v_{\mu'}, \qquad (1)$$

Here, $2|v_{\mu'}|/L$ represents the boundary scattering term, added to the diagonal elements of the scattering matrix $\Gamma_{\mu\mu'}$. The convergence of the $q$-point mesh for ALD-BTE calculations was systematically tested (Figure S8), resulting in the adoption of a 1×1×151 mesh for all N-SWCNT models. Figure 1b displays the $\kappa(L)$ for 1-SWCNT (red) and 7-SWCNT bundle (blue), calculated by ALD-BTE method using both BE and EQ statistics for comparative analysis. The results for all five studied systems are shown in Figure S9. For context, the SHC results from Figure 1a are replotted. When intrinsic phonon scattering in Eq. (1) is omitted, the system approaches the ballistic limit (dashed lines), and both $\kappa(L)_{\text{ALD-BTE(BE)}}$ and $\kappa(L)_{\text{ALD-BTE(EQ)}}$ increase linearly with length. In this ballistic regime, $\kappa(L)_{\text{ALD-BTE(EQ)}}$ consistently exceeds $\kappa(L)_{\text{ALD-BTE(BE)}}$ because the equipartition theorem assigns a higher heat capacity, $C_\mu(\text{EQ})$, to each mode than $C_\mu(\text{BE})$.[25, 28, 32] This trend persists at shorter lengths, where $\kappa_{\text{ALD-BTE(BE)}}$ remains systematically lower than $\kappa_{\text{SHC}}$, $\kappa_{\text{NEMD}}$, and $\kappa_{\text{ALD-BTE(EQ)}}$ (the latter three all employing EQ statistics). The influence of anharmonic scattering intensifies with system length,[27] leading to a crossover at ~500 nm for the 1-SWCNT, beyond which $\kappa_{\text{ALD-BTE(BE)}}$ exceeds $\kappa_{\text{ALD-BTE(EQ)}}$. This marks a transition to a regime where full phonon dispersion and detailed scattering physics become essential. The origin of this crossover lies in the distinct phonon populations governed by the different statistics (SI, Section S4.2). BE statistics ($\bar{n}_\mu^{\text{BE}} = 1/(e^{\hbar\omega/k_\text{B}T} - 1)$) exponentially suppress high-frequency mode occupation, while EQ statistics overpopulate these modes, as $\bar{n}_\mu^{\text{EQ}} = k_\text{B}T/(\hbar\omega)$, decay only inversely with frequency. This overpopulation in the classical case increases the available scattering channels, thereby reducing $\kappa_{\text{ALD-BTE(EQ)}}$.[27, 28] For the 7-SWCNT bundle, $\kappa_{\text{ALD-BTE(BE)}}$ and $\kappa_{\text{ALD-BTE(EQ)}}$ converge with only a small deviation, where $\kappa_{\text{ALD-BTE(EQ)}}$ is slightly higher. This indicates that the difference between quantum and classical statistical treatments diminishes in large bundles at longer lengths. The close agreement between $\kappa_{\text{SHC}}$ and $\kappa_{\text{ALD-BTE(EQ)}}$ for the isolated SWCNT validates the accuracy of the trained NEP. At extended lengths, $\kappa_{\text{SHC}}$ falls slightly below $\kappa_{\text{ALD-BTE(EQ)}}$, a well-understood consequence of the inclusion of higher-order anharmonic scattering in the SHC formalism.[33] Remarkably, upon bundling, the trend reverses. This switch indicates that the discrepancy cannot be attributed solely to differences in phonon statistics, but instead points to a deeper methodological divergence that emerges only in coupled nanotube systems.

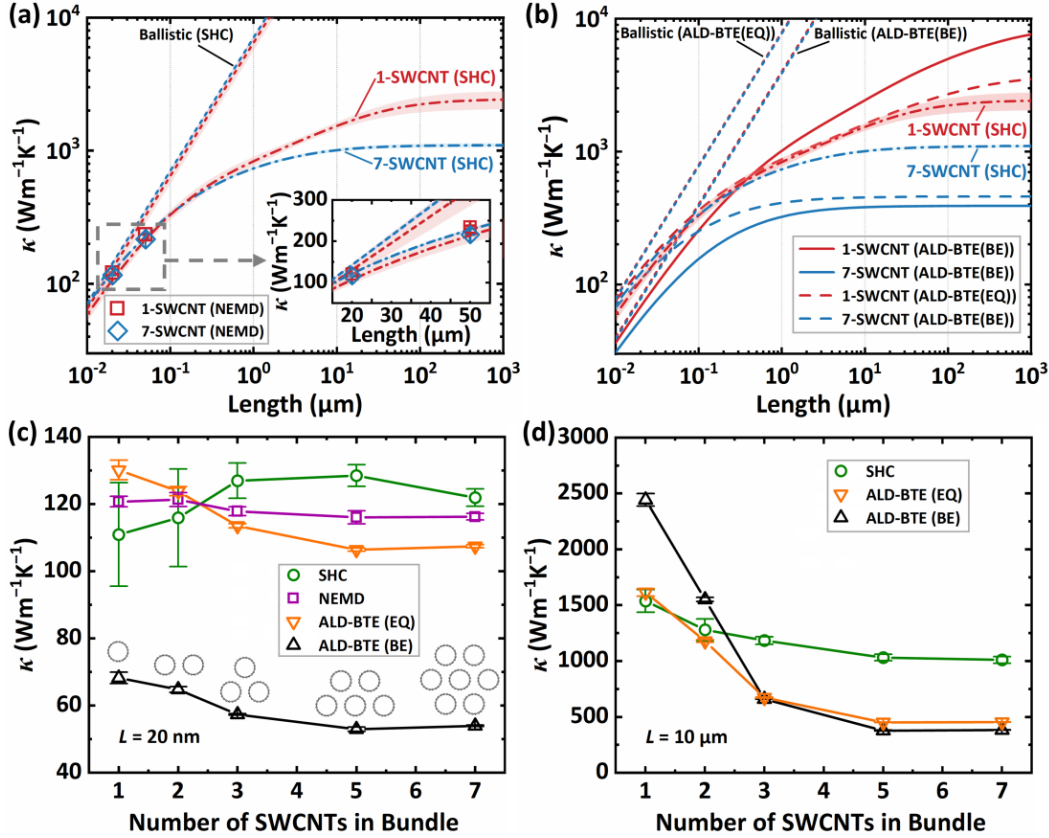

**Figure 1. Thermal transport in single and bundled SWCNTs at 300 K, calculated using the NEP.** (a, b) Length-dependent thermal conductivity $\kappa(L)$ for a 1-SWCNT (red) and a 7-SWCNT bundle (blue) from: (a) SHC method (dash-dot curves) and NEMD simulations (scattered points at 20 nm and 50 nm); (b) ALD-BTE using Bose-Einstein (BE, solid curves) and equipartition (EQ, dashed curves) statistics. In (a, b), dotted lines denote the corresponding ballistic limits. (c, d) $\kappa$ as a function of the number of SWCNTs (N) at lengths of (c) 20 nm and (d) 10 μm, from SHC (green circles), NEMD (purple squares), ALD-BTE (EQ) (orange inverted triangles), and ALD-BTE (BE) (black triangles). NEMD was not performed at 10 μm due to computational constraints. The schematic in (c) illustrates the geometric arrangement of the N-SWCNT bundles. Error representations: shaded regions in (a, b) and error bars in (c, d) denote standard deviations. For SHC and NEMD, these were obtained from three independent simulation runs; for ALD-BTE, they were calculated from three independent computations using a converged *q*-point mesh.

Having compared the $\kappa$ of single and bundled SWCNTs across different simulation methods, we now turn to the specific impact of bundle size. In the ballistic regime, both $\kappa(L)_{\text{ALD-BTE(BE)}}$ and $\kappa(L)_{\text{ALD-BTE(EQ)}}$ for bundles coincide with that of the corresponding single (1-)SWCNT. However, when phonon scattering is considered, the growth rates of $\kappa(L)_{\text{ALD-BTE(BE)}}$ and $\kappa(L)_{\text{ALD-BTE(EQ)}}$ with length are suppressed for all N-SWCNTs, with the suppression strength increasing with N. For instance, at $L = 10$ μm, the $\kappa_{\text{ALD-BTE(BE)}}$ of the 7-SWCNT bundle is 84 % lower than that of the 1-SWCNT (Figure 1d), consistent with experimental reports of substantial $\kappa$ reduction in bundles.[2,5] In the EQ statistics, a 69 % reduction is observed for the 7-SWCNT bundle. The smaller reduction compared to $\kappa_{\text{ALD-BTE(BE)}}$ stems from the lower $\kappa$ in 1-SWCNT by EQ statistics, while its impact on larger bundles is insignificant, as discussed above. Moreover, as illustrated in Figure 1c, $\kappa(L)_{\text{ALD-BTE(BE)}}$ and $\kappa(L)_{\text{ALD-BTE(EQ)}}$ exhibit a continuous decline with increasing bundle size even at short lengths ($L = 20$ nm). At this length, the 7-SWCNT bundle shows reductions of 16 % ($\kappa(L)_{\text{ALD-BTE(BE)}}$) and 17 % ($\kappa(L)_{\text{ALD-BTE(EQ)}}$) compared to an individual SWCNT. This stands in sharp contrast to the results from $\kappa(L)_{\text{SHC}}$ and $\kappa(L)_{\text{NEMD}}$, which show no systematic trend with bundle size, highlighting their limitations in capturing subtle inter-tube scattering effects in the quasi-ballistic regime. The ALD-BTE calculations indicate that the parallel circuit model remains valid only under ballistic conditions, beyond this regime, even at lengths as short as 20 nm, the model fails to describe thermal transport accurately. Notably, at 1 mm, the $\kappa_{\text{ALD-BTE(BE)}}$ of individual SWCNT approaches approximately 9000 Wm$^{-1}$K$^{-1}$ (Figure 1b), with its convergence length exceeding 1 mm, in accordance with previous experimental observations of $\kappa$ increasing continuously up to millimeter-scale lengths.[1]

To elucidate the intrinsic mechanisms behind the significant reduction in the thermal conductivity of SWCNT bundles, we systematically analyze key phonon modes for each N-SWCNT system. The phonon dispersion relations below 300 cm$^{-1}$ are shown in Figure 2a,b, where the color mapping represents the mode-resolved $\kappa$ and scattering rate $\Gamma$, respectively, as calculated from ALD-BTE(BE). The results reveal that the longitudinal acoustic (LA) and twisting (TW) modes exhibit low scattering rates and high group velocities, making them the dominant contributors to $\kappa$. The doubly degenerate transverse acoustic (TA) modes also provide a substantial contribution, though it is secondary to that of the LA/TW modes due to their lower group velocity. In contrast, optical modes contribute negligibly to $\kappa$, a consequence of their high $\Gamma$ and diminished group velocity.

Bundling increases phonon scattering rate, thereby suppressing the contributions of the LA, TW, and TA modes to $\kappa$. As quantified in Figure 2c, the suppression is most severe for the TW mode, which undergoes a 92.2 % reduction in a 2-SWCNT bundle and a near-complete 99.6 % reduction in a 3-SWCNT bundle (Figure 2c). This catastrophic decline stems from the breaking of rotational symmetry upon bundling, which replaces the rigid twisting motion of an isolated nanotube with localized liberational oscillations between adjacent SWCNTs.[34] This symmetry breaking is directly evidenced by the up-shift of the TW mode frequency from zero to a finite value (about 50 cm$^{-1}$) at the $\Gamma$-point in the 7-SWCNT bundle (Figure 2a), a finding consistent with previous report (16-50 cm$^{-1}$).[35] Consequently, the scattering rate $\Gamma$ of the TW mode increases more significantly than those of the LA and TA modes (Figure 2b), explaining the dramatic reduction in $\kappa$ observed during the transition from a single SWCNT to a 3-SWCNT bundle (Figure 1d). Beyond symmetry breaking, increasing the number of SWCNTs in the bundle induces a systematic proliferation of phonon modes across all frequency spectrum, progressively narrowing of both acoustic-optical and optical-optical phonon band gaps, and thereby increase in available phonon scattering channels. This causes a substantial expansion of the three-phonon scattering phase space (Figure

S11e) and a concomitant increase in scattering rates $\Gamma$ across the entire phonon spectrum (Figure 2b). Together, these effects produce strong reductions in the LA and TA modes, by approximately 80 % and 60 %, respectively, when one SWCNT is bundled with two others.

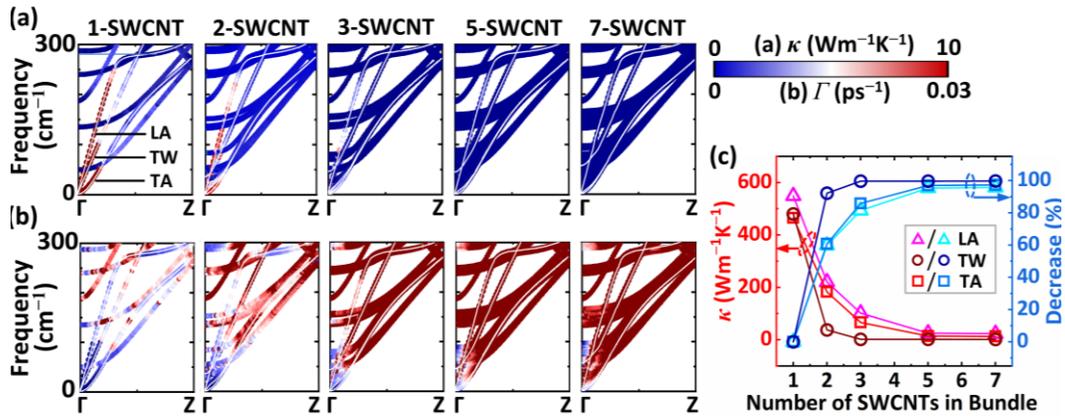

**Figure 2. Phonon dispersion and transport properties of SWCNT bundles.** Phonon dispersion relations for N-SWCNT systems below 300 cm$^{-1}$, colored by the (a) mode-resolved thermal conductivity ($\kappa$) and (b) scattering rate ($\Gamma$) for each phonon branch, as calculated from ALD-BTE with Bose-Einstein (BE) statistics. In (a) and (b), the longitudinal acoustic (LA), twisting (TW), and transverse acoustic (TA) branches are labeled, with white curves guiding the eye along each dispersion. (c) Thermal conductivity ($\kappa$) contributed by the LA, TW, and TA modes (left axis) and their reduction rate (right axis) as a function of bundle size.

To further elucidate phonon transport behavior in the simulated systems, we analyzed the key factors governing $\kappa$, as expressed in Eq. (1), which are heat capacity ($C_\mu$), phonon group velocity ($v_\mu$), mean free path (MFP, $\lambda_\mu$) and scattering rate ($\Gamma_\mu$). As compared in Figure 3a, the $C_\mu$ remains unchanged upon bundling into a 7-SWCNT structure. This invariance occurs because $C_\mu$ depends exclusively on vibrational frequency at a given temperature. While bundling introduces additional phonon modes, their heat capacity follows the same Bose-Einstein statistics as an isolated

SWCNT, resulting in identical frequency dependence. At room temperature (300 K), low-frequency phonons ($\hbar\omega < k_BT$, corresponding to $\omega < 208.5$ cm$^{-1}$, the orange shaded area) are thermally populated, with their heat capacity approaching the classical limit, $k_B$.[21] In contrast, high-frequency modes ($\hbar\omega > k_BT$, the green shaded area) remain largely unexcited,[36] leading to an exponential decay in $C_\mu$ with increasing frequency.

The phonon group velocity ($v_\mu = \partial\omega/\partial q$, where $q$ is the wave vector) was derived from the dispersion relation. As shown in Figure 3b, the $v_\mu$ of phonon modes in the 7-SWCNT bundle, corresponding to those in the isolated 1-SWCNT, show only a marginal reduction. The increased population of low-velocity modes in the bundle stems predominantly from newly emergent phonon modes introduced by inter-tube interactions, rather than a substantial decrease in the group velocities of the intrinsic SWCNT phonon modes. In stark contrast, the MFP ($\lambda_\mu$) of the 7-SWCNT bundle decreases significantly across all frequencies (Figure 3c), with reductions reaching up to two orders of magnitude in the very low-frequency range, indicative of strongly enhanced phonon scattering.

Correspondingly, Figure 3d reveals that three-phonon scattering rates increase by approximately two orders of magnitude across the entire frequency spectrum. According to Fermi's golden rule,[37] the scattering rate for a phonon mode $\mu$ is proportional to both the scattering phase space ($P_\mu$) and the squared scattering matrix element ($|V_{\mu\mu'\mu''}|^2$). Our analysis shows that the enhancement in scattering rates is primarily driven by a significant increase in $P_\mu$ (Figure 3e), which quantifies the number of available three-phonon scattering channels in the bundled system. Counterintuitively, the scattering matrix element $|V_{\mu\mu'\mu''}|$, which governs the scattering strength, exhibits a pronounced reduction upon bundling (Figure 3f). This indicates that the expansion of the phase space overwhelms the concurrent weakening of individual scattering strengths.

Residual anharmonicity analysis (Figure S12) confirms that the anharmonic force constants decrease as the number of SWCNTs increases, consistent with the observed reduction in $|V_{\mu\mu'\mu''}|^2$. Consequently, unlike the scattering phase space, which grows with bundle size, $|V_{\mu\mu'\mu''}|^2$ diminishes (Figure S11f), partially counteracting the phase space effects. This competition results in a saturation of the scattering rates rather than indefinite scaling, ultimately leading to a converged thermal conductivity in larger bundles. Additional comparisons for all N-SWCNT systems are provided in Figure S11.

Therefore, two key factors contribute to the increased scattering rates in bundles. First, symmetry breaking: the cylindrical rotational symmetry of individual SWCNTs is disrupted in bundles, significantly increasing scattering rates for rotational-symmetry-sensitive phonons, such as the TW mode. Second, increased scattering channels: the additional phonon modes in bundles provide more pathways that satisfy energy and momentum conservation, enhancing scattering rates across all frequencies.

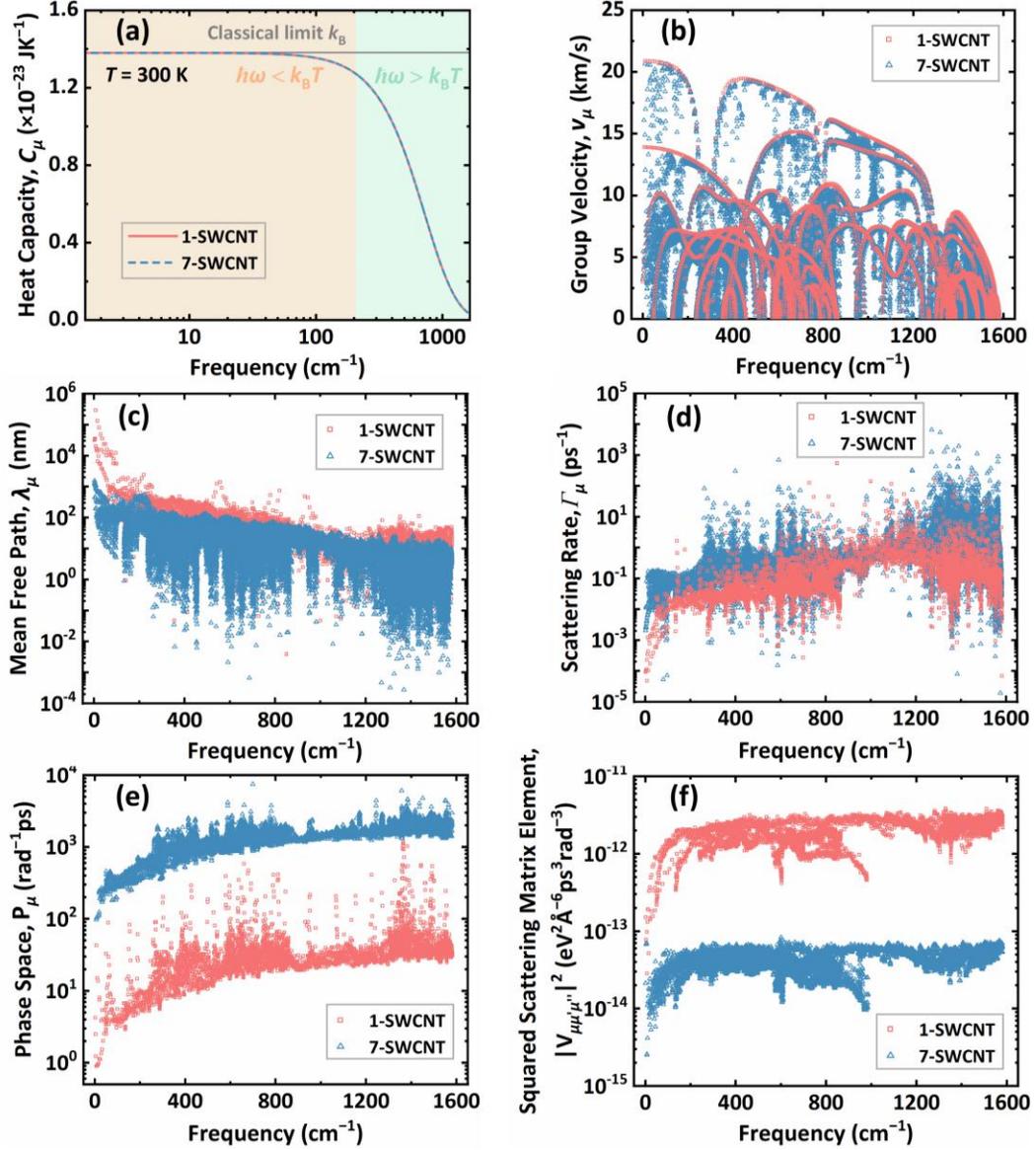

**Figure 3. Comparative phonon transport properties of an isolated (1-)SWCNT (red) and a 7-SWCNT bundle (blue).** (a) mode-resolved heat capacity $C_\mu$, (b) group velocity $v_\mu$, (c) MFP $\lambda_\mu$, (d) scattering rate $\Gamma_\mu$, (e) scattering phase space $P_\mu$, and (f) squared three-phonon scattering matrix element $|V_{\mu\mu'\mu''}|^2$. In (a), the gray solid line indicates the classical Dulong-Petit limit ($k_B$), while the orange and green shaded areas demarcate the low-frequency ($\hbar\omega < k_B T$) and high-frequency ($\hbar\omega > k_B T$) regimes at $T = 300$ K, with the transition at $\omega = 208.5$ cm$^{-1}$. All quantities were obtained using the ALD-BTE framework with Bose-Einstein (BE) statistics.

To understand the distinct bundling effects observed at short (20 nm) and long (10 μm) nanotube lengths (Figure 1c,d), we analyze the frequency-resolved thermal conductivity for both the isolated SWCNT and the 7-SWCNT bundle (Figure 4). For the 20 nm case, low-frequency phonon modes (with long MFP) are truncated by the system boundaries, while higher-frequency modes (> 208.5 cm$^{-1}$), with shorter intrinsic MFP, contribute approximately 82 % of the thermal conductivity. Notably, in the 7-SWCNT bundle, the contribution of high-frequency phonons is suppressed, reducing the overall thermal conductivity. At $L = 10$ μm, thermal conductivity is instead dominated by low-frequency phonons (< 208.5 cm$^{-1}$), accounting for approximately 65 % of the total, while high-frequency modes play a negligible role. Within this range, the reduction in thermal conductivity for the 7-SWCNT bundle primarily stems from the suppressed contributions of low-frequency phonons (Figure 4b). Additional information for all N-SWCNT are provided in Figure S13.

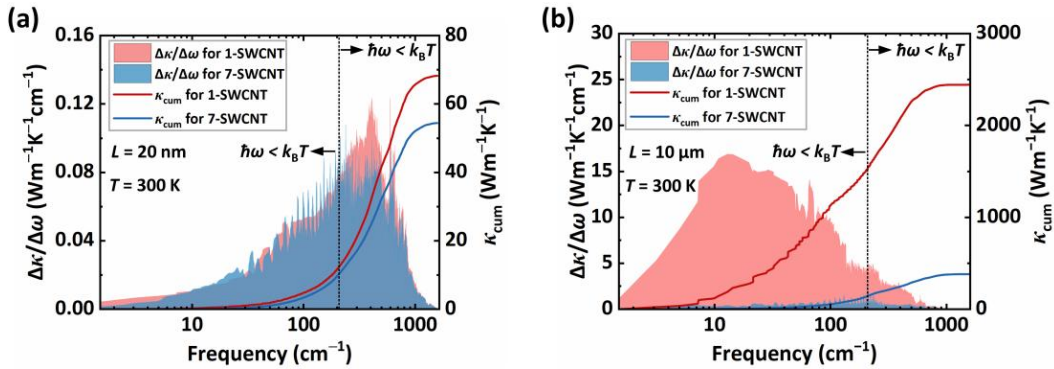

**Figure 4. Frequency-dependent thermal conductivity of isolated and bundled SWCNTs.** Spectral thermal conductivity (shaded areas, left axis) and its cumulative value (solid lines, right axis) for an isolated SWCNT (red) and a 7-SWCNT bundle (blue) at 300 K, calculated using the ALD-BTE framework with Bose-Einstein (BE) statistics. Results are shown for system lengths of (a) 20 nm and (b) 10 μm.

We now elucidate the origin of the crossover between $\kappa_{\text{ALD-BTE(BE)}}$ and $\kappa_{\text{ALD-BTE(EQ)}}$ with increasing bundle size, as observed in Figure 1d. To this end, we computed the MFP using both EQ and BE statistics within the ALD-BTE framework for 1-SWCNT and the bundled systems (Figure 5a and S14a–e). In the 1-SWCNT, the MFP derived from EQ statistics is significantly shorter than that from BE statistics. Although the heat capacity under the equipartition theorem, $C_\mu(\text{EQ})$, is larger than $C_\mu(\text{BE})$ (Figure S11a),[25] this does not compensate for the pronounced reduction in MFP, resulting in a substantially lower $\kappa_{\text{ALD-BTE(EQ)}}$ compared to $\kappa_{\text{ALD-BTE(BE)}}$. In contrast, for the 7-SWCNT bundle, the two statistical treatments yield nearly identical MFP, with EQ statistics only predicts marginally lower MFP in the mid- to high-frequency range. Consequently, the larger $C_\mu(\text{EQ})$ becomes the dominant factor, leading to a slightly higher $\kappa_{\text{ALD-BTE(EQ)}}$ than $\kappa_{\text{ALD-BTE(BE)}}$ (Figure 1d). This trend is systematic: the discrepancy in MFP between EQ and BE statistics gradually decreases with increasing bundle size (Figure S14a–e). Since the MFP is inversely proportional to the scattering rate, this convergence indicates that the inherently higher three-phonon scattering rates predicted by EQ statistics are strongly suppressed in bundled configurations. This suppression narrows the gap between $\kappa_{\text{ALD-BTE(BE)}}$ and $\kappa_{\text{ALD-BTE(EQ)}}$ with increasing bundle size, explaining the initial convergence and subsequent crossover shown in Figure 1d.

The discrepancies in scattering rates between EQ and BE statistics originates from their respective treatments of phonon population, $\bar{n}_\mu^{\text{EQ}}$ and $\bar{n}_\mu^{\text{BE}}$, which weight the three-phonon scattering processes.[38] These processes are categorized into absorption and emission events.[39] In an absorption process, a phonon mode $\mu$ combines with another mode $\mu'$ to create a third mode $\mu''$. Conversely, in an emission process, a phonon mode $\mu$ decays into two new modes $\mu'$ and $\mu''$. The weighting factors differ between the two statistical frameworks: under EQ statistics, the factors are $\bar{n}_{\mu'}^{EQ} - \bar{n}_{\mu''}^{EQ}$ for absorption and $\bar{n}_{\mu'}^{EQ} + \bar{n}_{\mu''}^{EQ}$ for emission; under BE statistics, the corresponding factors are $\bar{n}_{\mu'}^{BE} - \bar{n}_{\mu''}^{BE}$ for absorption and $1 + \bar{n}_{\mu'}^{BE} + \bar{n}_{\mu''}^{BE}$ for emission.[38, 40] To quantify the difference in how these statistics weight the scattering processes for a given phonon mode $\mu$, we define a dimensionless ratio $R_\mu$ as the weighting factors under EQ statistics divided by that under BE statistics. For absorption and emission processes, $R_\mu$ is defined respectively as:

$$R_\mu^{abs} = \frac{\sum_{\mu',\mu''} \bar{n}_{\mu'}^{EQ} - \bar{n}_{\mu''}^{EQ}}{\sum_{\mu',\mu''} \bar{n}_{\mu'}^{BE} - \bar{n}_{\mu''}^{BE}}, \tag{2}$$

$$R_\mu^{emi} = \frac{\sum_{\mu',\mu''} \bar{n}_{\mu'}^{EQ} + \bar{n}_{\mu''}^{EQ}}{\sum_{\mu',\mu''} 1 + \bar{n}_{\mu'}^{BE} + \bar{n}_{\mu''}^{BE}}, \tag{3}$$

A deviation factor $R_\mu > 1$ indicates a larger weight for the scattering processes for phonon mode $\mu$ from EQ statistics, resulting in a larger number of phonon scattering processes and a shorter MFP. Conversely, $R_\mu < 1$ signifies a smaller weight for the scattering processes from EQ statistics, leading to a lower number of phonon scattering processes and a longer MFP. As shown in Figure 5b for both the 1-SWCNT and the 7-SWCNT bundle, the $R_\mu^{abs}$ consistently exceeds 1, while the $R_\mu^{emi}$ is always less than 1. This confirms that the significantly shorter MFPs predicted by EQ statistics in the isolated SWCNT, and for the mid- to high-frequency modes in the bundle (Figure 5a), originate predominantly from the artificially enhanced number of phonon absorption processes inherent to the EQ approximation. Notably, $R_\mu^{abs}$ exhibits a pronounced reduction across all frequencies in the 7-SWCNT bundle compared to the 1-SWCNT. Below 22 cm$^{-1}$, it almost approaches unity ($R_\mu = 1$), indicating that the MFP predicted by EQ and BE statistics converge. In contrast, $R_\mu^{emi}$ decreases only slightly above 900 cm$^{-1}$, confirming that emission processes play a negligible role, as their influence is both small in magnitude and restricted to high frequencies. Furthermore, the discrepancy between EQ- and BE-derived MFPs systematically diminishes with increasing bundle size, as corroborated by Figure S14. This trend, combined with the consistently larger heat capacity per mode $C_\mu$ calculated with EQ statistics compared to BE statistics, collectively explains the divergence between $\kappa_{ALD-BTE(BE)}$ and $\kappa_{ALD-BTE(EQ)}$ observed in Figure 1d.

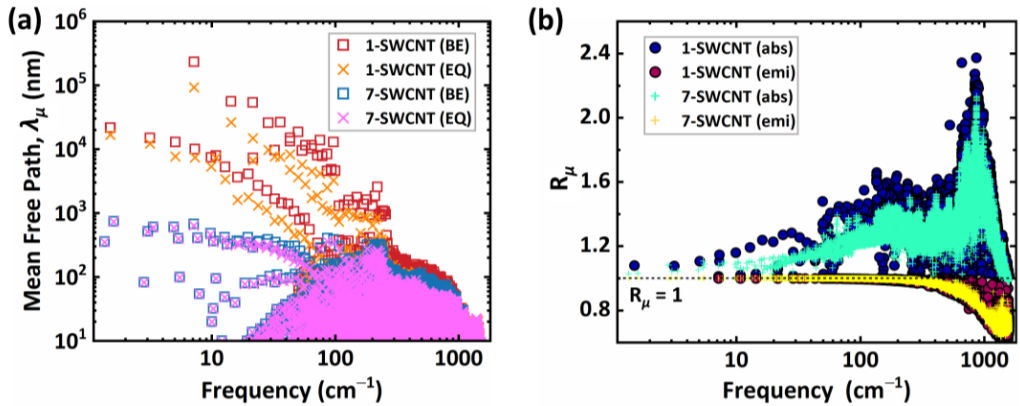

**Figure 5. Influence of phonon distribution statistics on thermal transport in single and bundled SWCNTs.** Comparison of (a) the phonon MFP $\lambda_\mu$, and (b) weighting factors ratio $R_\mu$, for three-phonon scattering processes. In (b), the dashed black line ($R_\mu = 1$) represents the boundary where scattering is equally probable under both statistical descriptions.

The agreement between our ALD-BTE calculations and previous experimental measurements[2,5] for the thermal conductivity $\kappa$ of SWCNT bundles (*N*) at $L$ = 5 μm and 300 K is shown in Figure 6. Our model accurately captures the significant reduction in $\kappa$ with increasing *N*, predicting an 81 % decrease for a 7-SWCNT bundle that is in excellent agreement with measurement. This close correspondence provides strong validation for the predictive power of our computational framework. It is worth noting that for N > 5, the calculated $\kappa$ converges to a size-independent value, whereas the experimental measurements continue to decrease, albeit at a decreasing rate. This emerging discrepancy for larger bundles can be attributed to two primary factors. First, due to the substantial computational cost of modeling large systems, our ALD-BTE calculations are limited to three-phonon scattering processes; the potential suppression of $\kappa$ by higher-order phonon scatterings (*e.g.*, four-phonon) in large bundles remains unaccounted for. Second, our model assumes an ideal, parallel alignment of nanotubes. In contrast, experimentally synthesized bundles often exhibit increased structural disorder, such as tube entanglements and misorientations, with increasing bundle size, which introduces additional phonon scattering channels and further reduces the measured thermal conductivity.[11]

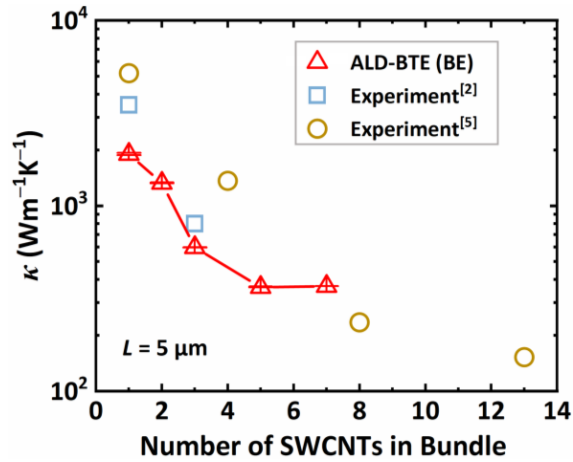

**Figure 6. Thermal conductivity reduction in SWCNT bundles.** Comparison of the thermal conductivity ($L$ = 5 μm, $T$ = 300 K) as a function of bundle size ($N$) between our ALD-BTE calculations with Bose-Einstein (BE) statistics (red triangles) and experimental data sets (blue squares, brown circles). The ALD-BTE(BE) results quantitatively capture the experimentally observed suppression of thermal conductivity with increasing bundle size. Error bars on the ALD-BTE data represent standard deviations from three independent calculations using a converged *q*-point mesh.

## CONCLUSIONS

In summary, by developing a NEP for SWCNTs and integrating it with the ALD-BTE framework, we achieve a quantitative, mode-resolved understanding of thermal transport from individual (10, 0) zigzag SWCNT to their bundled assemblies. Our approach successfully reproduces experimental observations, predicting a pronounced 81 % reduction in thermal conductivity for a 5-μm-long bundle of seven SWCNTs. Notably, even at a length of 20 nm, an 18 % reduction is observed, definitively disproving the validity of the parallel-circuit model in quasi-ballistic regimes. Phonon-mode analysis reveals that bundling breaks the rotational symmetry of SWCNTs, which strongly suppresses symmetry-sensitive phonon modes and concurrently elevates three-phonon scattering rates across the entire spectrum, collectively leading to a dramatic suppression of thermal conductivity. Critically, we demonstrate that accurate modeling necessitates quantum statistics; the use of classical equipartition principles directly undermines agreement with experiment. This work not only elucidates the microscopic mechanisms governing thermal transport in SWCNT bundles but also provides essential theoretical guidance for the rational design of SWCNT-based bulk materials for macroscopic thermal-management applications.

## METHODS

To minimize excessive axial strain, we considered bundles composed of SWCNTs with uniform chirality indices.[12] Since chirality has a negligible effect on the thermal conductivity of SWCNTs,[41] we constructed models of N-SWCNT bundles using (10, 0) zigzag SWCNTs as a representative case.

We trained a high-accuracy NEP using the GPUMD package.[42] The training dataset construction is detailed in Section S1.1 of the SI, with model parameters and validation data provided in Section S1.2.

Employing this potential, we systematically investigated the thermal conductivity of SWCNTs using both MD simulations and the ALD-BTE framework, with full methodological details in the Sections S2.1 and S4.1. For MD simulations, we employed the Sääskilahti-Chalopin SHC method,[29,43] which offers a computationally efficient alternative to NEMD for evaluating length-dependent thermal conductivity, as it avoids NEMD's strong system-size dependence.

Within the ALD-BTE framework, thermal conductivity is computed by solving the linearized BTE for the nonequilibrium phonon population, $\delta n_\mu^{BE} = n_\mu^{BE} - \bar{n}_\mu^{BE}$, where the $\bar{n}_\mu^{BE}$ is the equilibrium BE distribution.[44,45] To isolate the specific role of quantum statistics, we developed a classical analogue (ALD-BTE(EQ)) by substituting $\bar{n}_\mu^{BE}$ with its classical counterpart, $\bar{n}_\mu^{EQ} = k_B T/(\hbar \omega)$. Notably, the relaxation time approximation fails here due to prominent hydrodynamic effects.[46,47] Moreover, iterative self-consistent solutions converge only when the off-diagonal elements of the scattering tensor are smaller in magnitude than the diagonal ones ($|\Gamma_{\mu\mu'}/\Gamma_{\mu\mu}| < 1$ for $\mu \neq \mu'$).[48] This criterion is violated for low-frequency phonons in the (10, 0) SWCNT system,[28] necessitating direct inversion of the scattering tensor to obtain an exact solution to the BTE.

## ASSOCIATED CONTENT

The Supporting Information is available free of charge.

The training dataset construction, model parameters, NEP validation, details of the MD simulation, SHC method verification, comparison of Tersoff and NEP, ALD-BTE computational details, convergence tests, phonon population under BE and EQ statistics, and phonon transport properties for all N-SWCNT systems.

## AUTHOR INFORMATION

**Corresponding Authors**

Ya Feng − Key Laboratory of Ocean Energy Utilization and Energy Conservation of Ministry of Education, School of Energy and Power Engineering, Dalian University of Technology, Dalian 116024, China; Department of Mechanical Engineering, The University of Tokyo, Tokyo 113-8656, Japan. Email: fengya@dlut.edu.cn

Shigeo Maruyama − State Key Laboratory of Fluid Power and Mechatronic Systems, School of Mechanical Engineering, Zhejiang University, Hangzhou 310027, China; Department of Mechanical Engineering, The University of Tokyo, Tokyo 113-8656, Japan; Institute of Materials Innovation, Institutes for Future Society, Nagoya University, Nagoya 464-8603, Japan. Email: maruyama@photon.t.u-tokyo.ac.jp

**Authors**

Feng Tao − Key Laboratory of Ocean Energy Utilization and Energy Conservation of Ministry of Education, School of Energy and Power Engineering, Dalian University of Technology, Dalian 116024, China;

Xiaoliang Zhang − Key Laboratory of Ocean Energy Utilization and Energy Conservation of Ministry of Education, School of Energy and Power Engineering, Dalian University of Technology, Dalian 116024, China;

Dawei Tang − Key Laboratory of Ocean Energy Utilization and Energy Conservation of Ministry of Education, School of Energy and Power Engineering, Dalian University of Technology, Dalian 116024, China;

**Author Contributions**

F. T.: investigation, methodology, writing – original draft, writing – review & editing; S. M.: formal analysis, supervision; Y. F.: conceptualization, data curation, formal analysis, funding acquisition, project administration, resources, supervision, writing – review & editing. All of the authors participated in the discussion and analysis of the work and commented on the manuscript.

**Notes**

The authors declare no competing financial interest.

ACKNOWLEDGMENT

Part of this work was supported by the National Natural Science Foundation of China (No. 524766164), LiaoNing Revitalization Talents Program (No. XLYC2403036), Science and Technology Innovation Fund of Dalian, International Science and Technology Cooperation (No. 2024JJ12RC035), the Fundamental Research Funds for the Central Universities (No. DUT23RC(3)053), and the National Key R&D Program of China (No. 2024YFA1409600). Part of this work was also supported by JSPS KAKENHI Grant Number JP23H00174, JP23H05443, JP21KK0087 and by JST, CREST Grant Number JPMJCR20B5, Japan.

# For Table of Contents Only

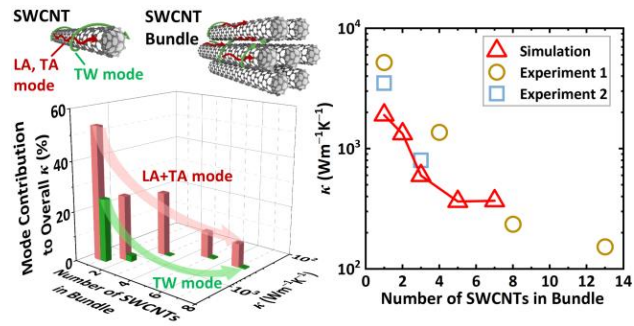